\documentclass{svjour3}                     % onecolumn (standard format)
\smartqed  % flush right qed marks, e.g. at end of proof
\usepackage{graphicx}
\begin{document}

\title{Numerical Studies of Counterflow Turbulence%\thanks{Grants or other notes
}
\subtitle{Velocity Distribution of Vortices}


\author{Hiroyuki Adachi         \and
        Makoto Tsubota %etc.
}


\institute{H.Adachi , M.Tsubota \at
              Department of Physics, Osaka City University, Sumiyoshi-Ku,
 Osaka 558-885 \\
              Tel.: +81-6-6605-3073\\
              Fax: +81-6-6605-2522\\
              \email{adachi@sci.osaka-cu.ac.jp}           %  \\
}

\date{Received: date / Accepted: date}

\maketitle

\begin{abstract}
We performed the numerical simulation of quantum turbulence produced by thermal counterflow in superfluid $^{4}${\rm He} by using the vortex filament model. The pioneering work was made by Schwarz, which has two defects. One is neglecting non-local terms of the Biot-Savart integral (localized induction approximation, LIA), and the other is the unphysical mixing procedure in order to sustain the statistically steady state of turbulence. We succeeded in making the statistically steady state without the LIA and the mixing. This state shows the characteristic relation $L=\gamma^2 v_{ns}^2$ between the line-length-density $L$ and the counterflow relative velocity $v_{ns}$ with the quantitative agreement of the coefficient $\gamma$ with some typical observations. We compare our numerical results to the observation of experiment by Paoletti {\it et al}, where thermal couterflow was visualized by solid hydrogen particles.

\end{abstract}

\section{Introduction}
\label{intro}
Quantum turbulence(QT), which consists of a tangle of quantized vortices, has
been investigated since the thermal counterflow experiments of Vinen [1-4]
 half a century ago, while the underlying physics
 is far from being fully understood [5,6]. 
The numerical simulations are the useful source of knowledge about QT, 
because the whole dynamics of this system is too complicated to be described
analytically. One of the powerful schemes of the simulation is the vortex
 filament model based on pioneering works by Schwarz [7,8].
Schwarz performed the numerical simulation of counterflow turbulence under the
periodic boundary condition by using
the localized induction approximation (LIA) which neglects a
non-local term of the Biot-Savart integral [8]. However he could not obtain
 the statistically steady state (SSS) because the vortices lie in
planes normal to ${\bf v}_{ns}={\bf v}_n-{\bf v}_s$ to prevent them from reconnecting.
 Therefore the unphysical mixing procedure, in which half of the vortices
are randomly selected to be rotated by 90$^\circ$ around the axis defined by 
the flow velocity, was used, and only this method enabled him to obtain the SSS.
 This failure reminds us that the LIA is unsuitable
 through the absence of the interaction between vortices.
Recently, thermal counterflow in superfluid $^4$He was visualized by using solid hydrogen particles, and the velocity distribution of particles was observed [9].
In this experiment, two types of particles appeared. Some particles move straight along the normalflow with the approximately same velocity as normalflow.
  Other particles move zigzag along the direction of superfluid and has different velocity from superfluid. In the latter case, particles seem to be trapped in the core of the vortices.
We try to apply our numerical results for understanding these observations.

Section 2 describes the equation of motion of vortices and the method of
numerical calculation. In Sec.3 we show the typical numerical results of
vortex tangle. Section 4 studies the validity of the LIA by comparing the LIA
 calculation with the full Biot-Savart calculation.
In the section 5 we present the velocity distribution of 
vortices in the SSS of vortex tangle and compare it with that of particles
 in the experimental observation.

\section{Equations of Motion and Numerical Simulation}
\label{sec:1}
\label{sec:2}
At 0K the velocity ${\dot{\bf s}}_0$ of the filament at the point {\bf s}($\xi$,t) is given by
\begin{equation}
{\dot {\bf s}}_0=\frac{\kappa}{4\pi}{\bf s}'\times{\bf s}'' \ln \biggl( \frac{2(l_+l_-)^{1/2}}{e^{1/4}a_0} \biggr)+\frac{\kappa}{4\pi}\displaystyle\int^{'}_{\it L}\frac{({\bf s}_1-{\bf s})\times d{\bf s}_1}{|{\bf s}_1-{\bf s|^3}}+{\bf v}_{s,a},
\end{equation}
where the prime denotes the derivatives with respect to the arc length $\xi$, and $\kappa$ is the quantized circulation, $a_0$ is a cutoff parameter corresponding
to a vortex core radius,
 $l_+$ and $l_-$ are the length of the two adjacent line elements that hold the
point{\bf s} between. The first term shows the localized induction field
 arising from a curved line element acting on itself. The second term 
represents the non-local field obtained by carrying out the integral of the
Biot-Savart integral along the rest of the filament. The third term $v_{s,a}$ is an applied field.
The LIA used in some works (e.g.,[7,8,10,11]) means neglecting the second
 non-local term. In contrast the calculation without the LIA is called the full
 Biot-Savart calculation.
At finite temperatures, the velocity ${\dot{\bf s}}$ is given by
\begin{equation}
{\dot{\bf s}}={\dot{\bf s}_0}+\alpha {\bf s}'\times ({\bf v}_n-{\dot{\bf s}}_0)-\alpha '{\bf s}'\times[{\bf s}'\times({\bf v}_n-{\dot{\bf s}}_0)],
\end{equation}
where $\alpha$ and $\alpha'$ are the temperature-dependent friction coefficients, ${\bf v}_n$ is the normalfluid velocity, and ${\dot{\bf s}_0}$ is calculated
 from Eq.(1). In the work [8] the author neglected the third term of Eq.(2).
We performed the full Biot-Savart calculation in this work.
The concrete reconnection procedure used in this work is the following.
Every vortex is represented by a string of points at intervals of almost 
$\delta \xi$. When a point on a vortex approaches another point on 
another vortex more closely than the fixed space resolution $\Delta \xi$,
we join these two points and reconnect the vortices [12].
%Since we apply the periodic boundary condition,
%for calculation the non-local term of eq,(2) we allocate
%26 boxes, which have the same configuration of vortices of the center main box,
%like a Rubik's Cube (Fig.1).
%        \begin{figure}[h]
%          \begin{minipage}{0.5\hsize}
%            \caption{Boxes allocated for calculation of a non-local
%                   contribution of vortices.
%                 The dark central box is the main computational box.}
%          \end{minipage}
%          \begin{minipage}{0.5\hsize}
%           \begin{center}
%            \scalebox{0.2}{\includegraphics{cubes.eps}}
%            \label{fig:cubes}
%           \end{center}
%          \end{minipage}
%        \end{figure}
For the integration of the motion of Eq.(2)
in time we used the classical 4th order Runnge-Kutta method.
 We usually start with an initial vortex configuration
of six vortex rings as shown in Fig.1(upper left).

\section{Simulation of Counterflow Turbulence}
In this section we present the numerical simulations of counterflow turbulence
at the temperature $T$=1.9K, in the computing box 0.1 $\times$ 0.1 $\times$ 0.1 ${\rm cm^3}$, at applied conterflow velocity $v_{ns}$=0.286, 0.381, 0.572 cm/s.
A typical result is shown in Fig.1.
\begin{figure}[h]
  \begin{center}
    \begin{tabular}{cc}
      \resizebox{40mm}{!}{\includegraphics{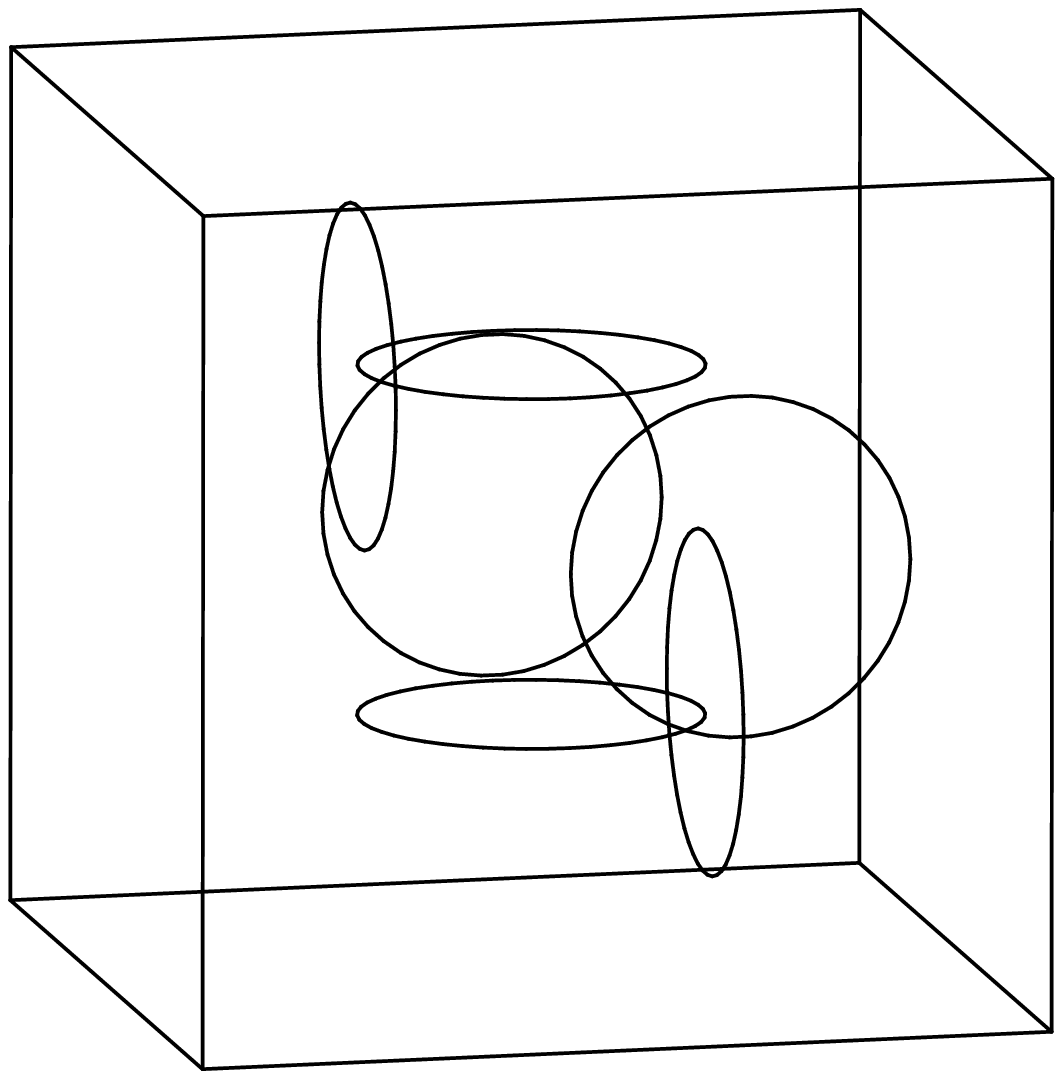}} &
      \resizebox{40mm}{!}{\includegraphics{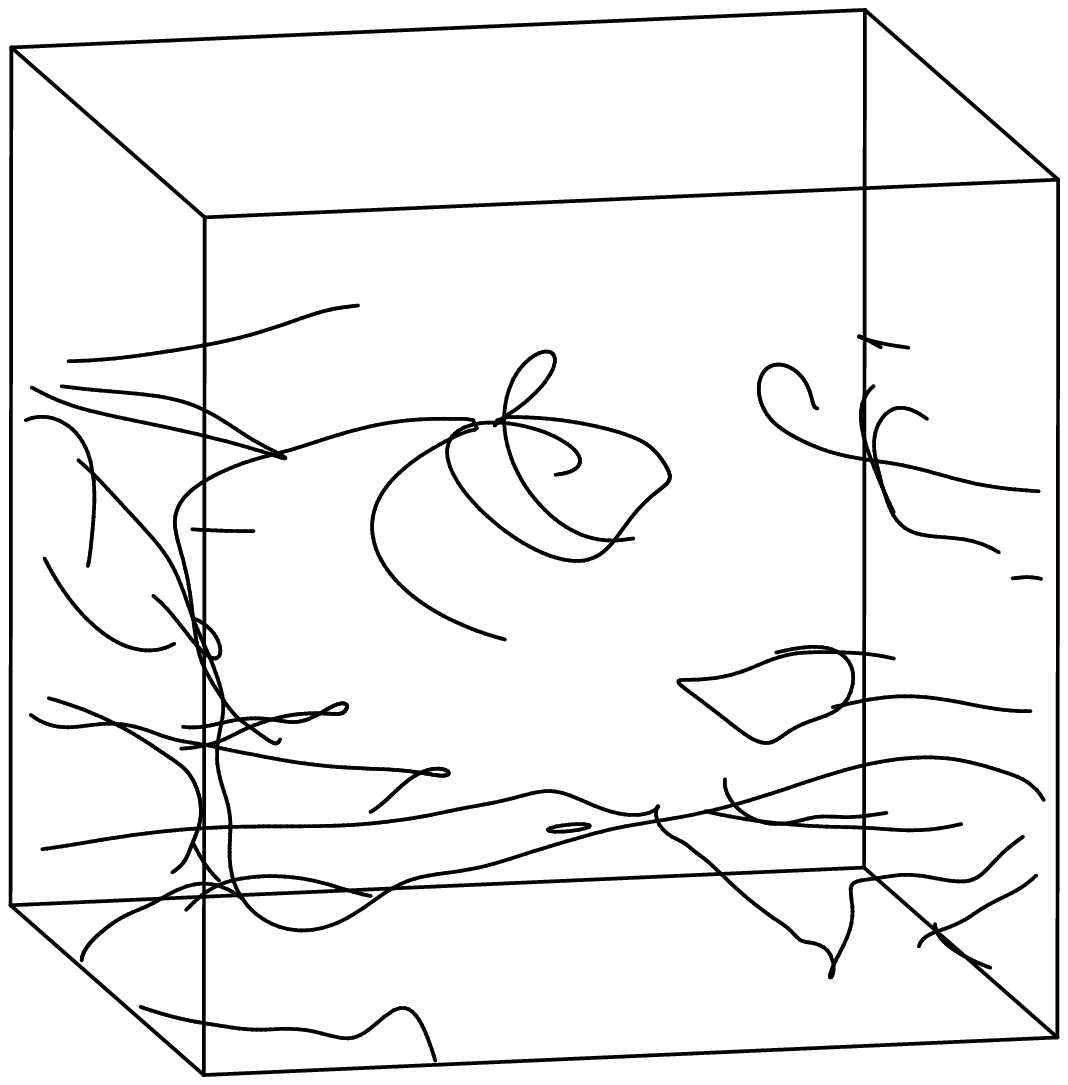}} \\
      \resizebox{40mm}{!}{\includegraphics{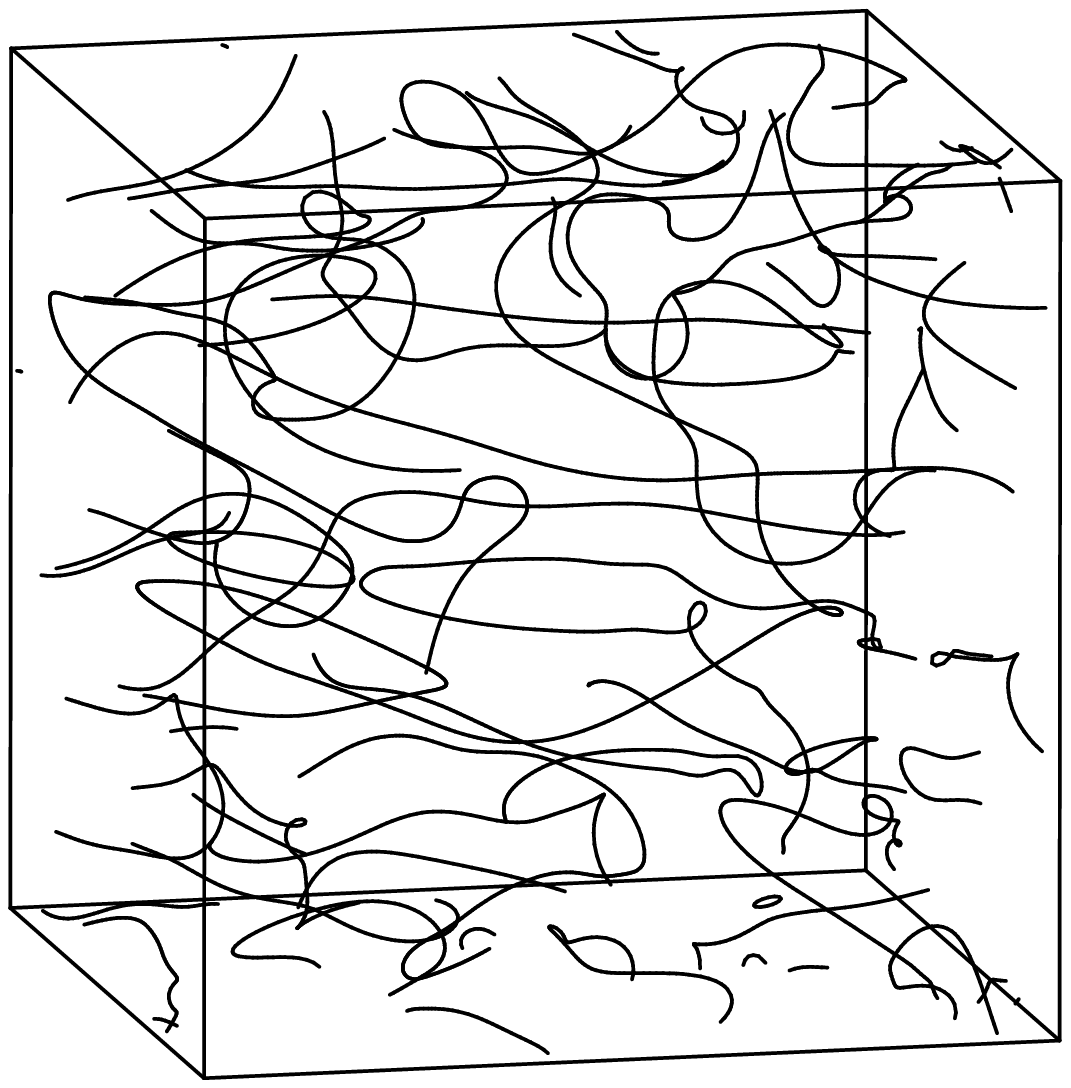}} &
      \resizebox{40mm}{!}{\includegraphics{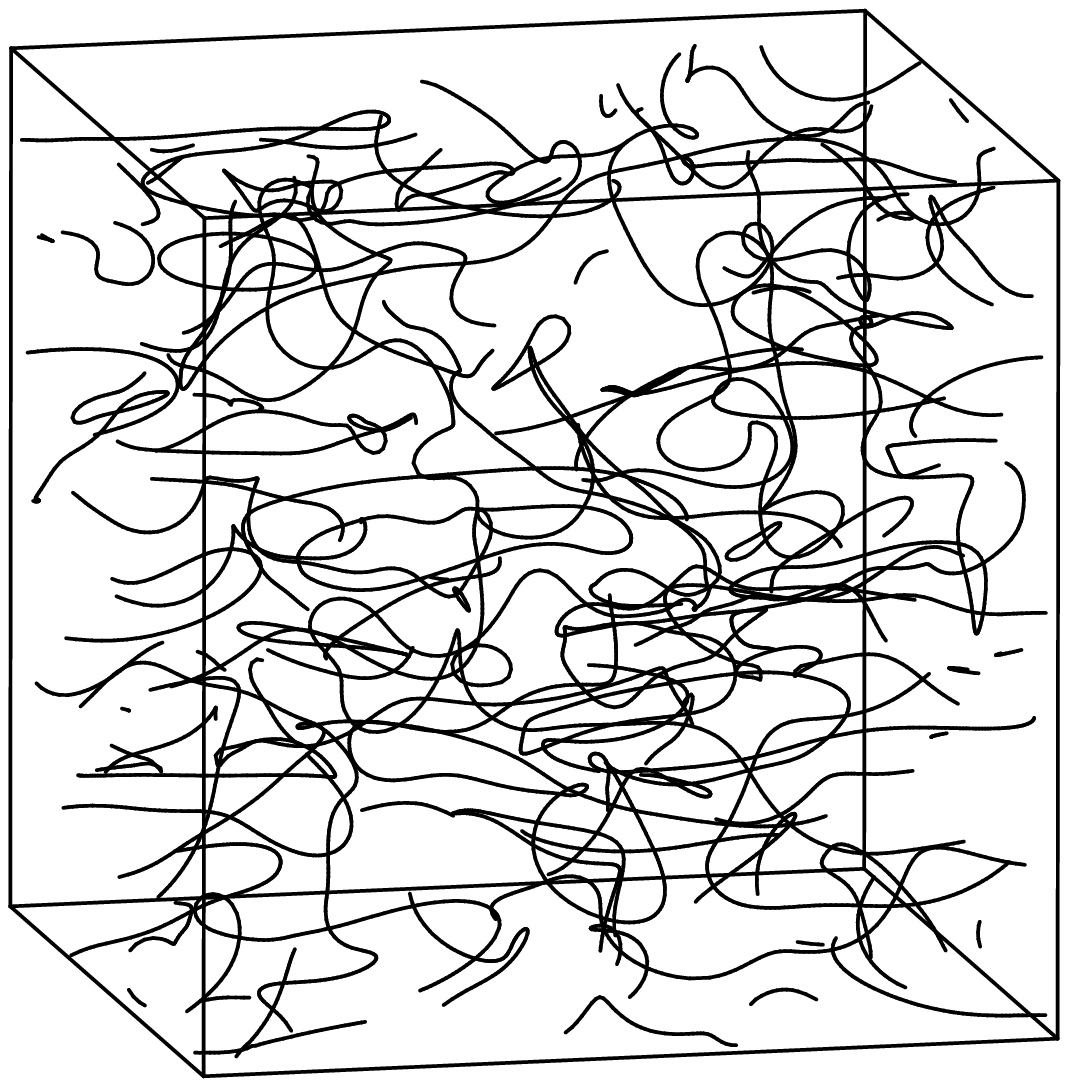}} \\
    \end{tabular}
     \begin{center}
      \caption{Development of a vortex tangle by the full Biot-Savart calculation in a periodical box with the size 0.1 cm. 
Here temperature $T$=1.9K, and conterflow velocity 
$v_{ns}$=0.572cm/s is along the vertical axis.
Upper left t=0 s, upper right t=0.5 s, lower left t=1.2 s, lower right t=2.5 s}
     \end{center}
    \label{test4}
  \end{center}
\end{figure}
Here the initial configuration of vortex loops evolves in the periodical box 
to a highly chaotic vortex tangle.
The vortex line density $L(t)$ is defined as the vortex line length per unit volume.
In Fig.2 we depict the time evolution of quantity $L(t)$. 
%        \begin{figure}[h]
%          \begin{minipage}{0.5\hsize}
%            \begin{center}
%             \scalebox{0.22}{\includegraphics{T19linedensity.eps}}
%             \label{fig:3a}             
%            \end{center}
%          \end{minipage}
%          \begin{minipage}{0.5\hsize}
%           \begin{center}
%            \scalebox{0.22}{\includegraphics{T19vinen.eps}}
%            \label{fig:3b}
%           \end{center}
%          \end{minipage}
\begin{figure}[h]
  \begin{center}
    \begin{tabular}{cc}
      \resizebox{52mm}{!}{\includegraphics{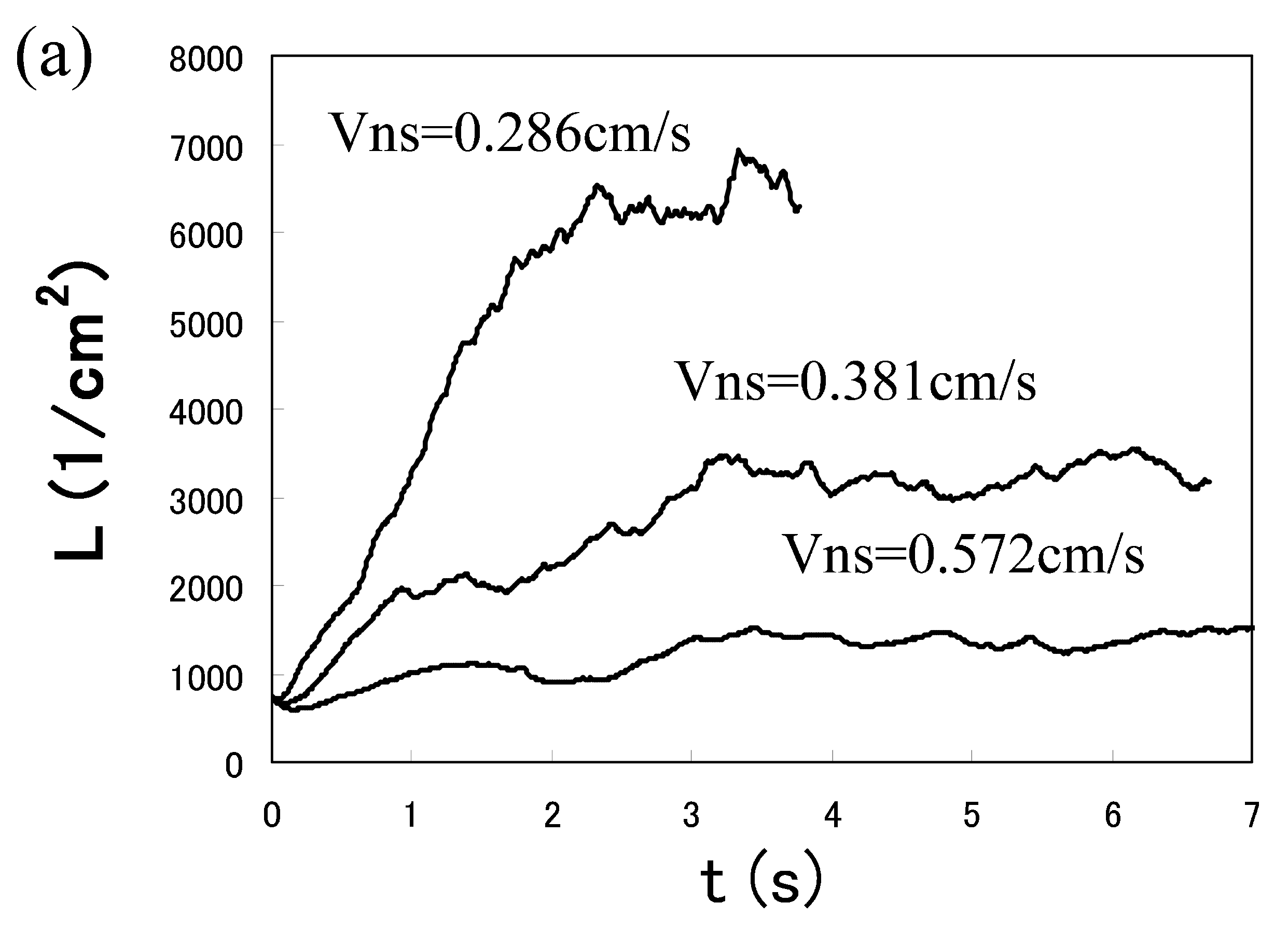}} &
      \resizebox{52mm}{!}{\includegraphics{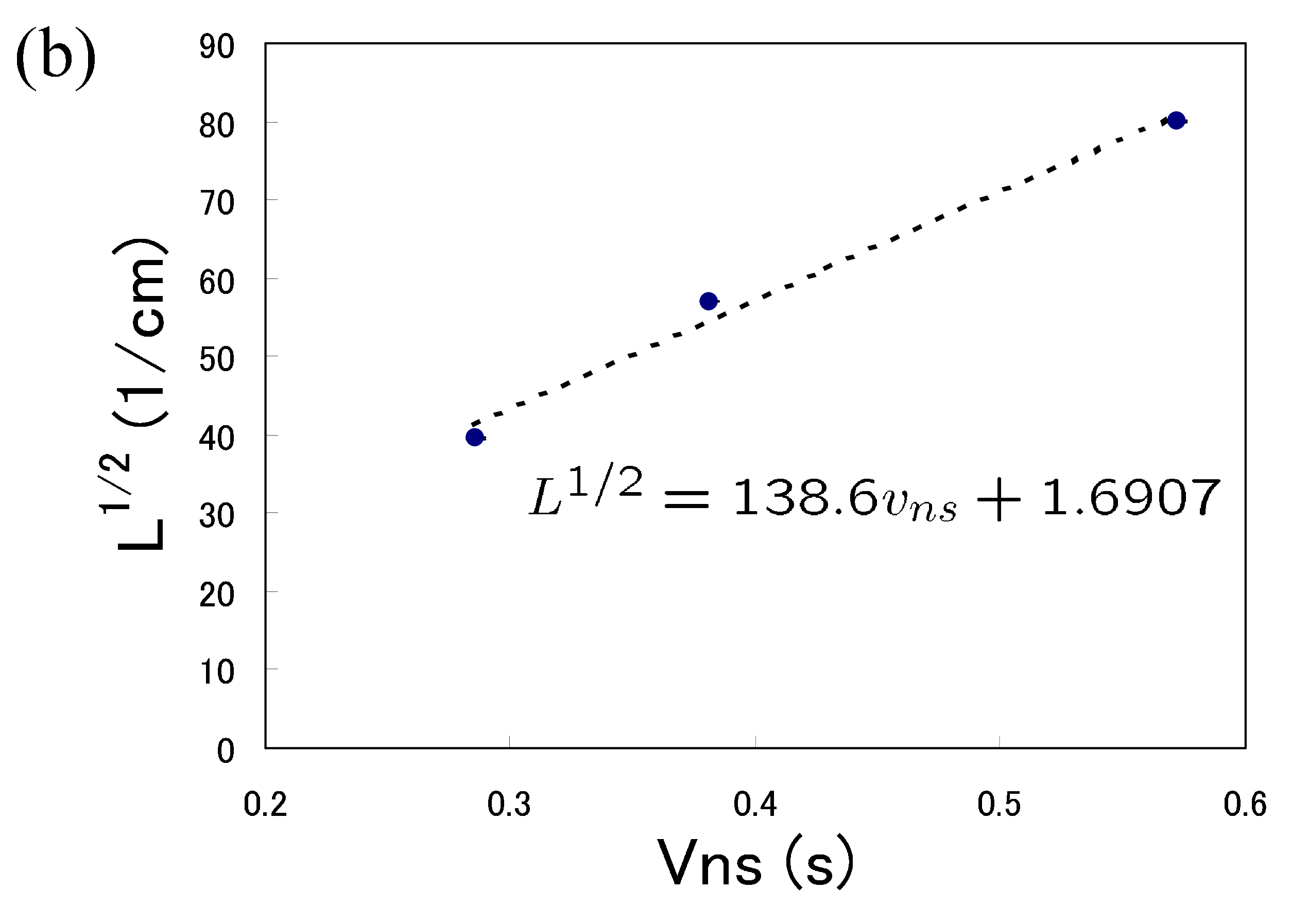}}
    \end{tabular}
     \begin{center}
      \caption{Vortex line density as a function of time for different
driving normalfruid ${\bf v}_{n}$} equal to 0.286, 0.381, 0.572 cm/s
     \end{center}
    \label{test4}
  \end{center}
\end{figure}
It is seen that
 the vortex tangle goes to the SSS after growth period.
In Fig.2(b) we found out that line length density satisfied 
the characteristic relation
 $L=\gamma^2 v_{ns}^2$ which have been obtained in previous experiments [13].
 This relation is derived from the Vinen's equation [3], and also obtained by
Schwarz using the LIA and the dynamical scaling [8].
We obtain $\gamma \approx 139{\rm s/cm^2}$ which quantitatively agrees with the experimental observation $\gamma \approx 130{\rm s/{\rm cm^2}}$ [14]. 

\section{Validity of the LIA}
In the similar works [8,10] the authors did not obtain the SSS of
turbulence (the work [8] created the SSS only with the mixing.)
The SSS was realized in the work [11] by using the LIA, in which the authors
mentioned that failures of previous works were due to the unsuitable 
reconnection procedure.
We will discuss the main reason why the SSS was not realized in previous works.
In order to consider the validity of the LIA, we compare two calculations.
One uses the LIA [Fig.3 left], and the other uses the full Biot-Savart
 law [Fig.3 right].
%        \begin{figure}[h]
%          \begin{minipage}{0.5\hsize}
%            \begin{center}
%            \scalebox{0.4}{\includegraphics{NBST16vn03t40.eps}}
%             \label{fig:4a}             
%            \end{center}
%          \end{minipage}
%          \begin{minipage}{0.5\hsize}
%           \begin{center}
%            \scalebox{0.4}{\includegraphics{BST16vn03t40.eps}}
%            \label{fig:4b}
%           \end{center}
%          \end{minipage}
%          \caption{Side view of vortex configuration by the LIA calculation
%(left) and by the full Biot-Savart law (right) at t=34.5s. The system is a $(0.2{\%rm cm})^3$ cube. Applied normalfluid velocity $v_{ns}=0.367{\rm cm/s}$.} 
%        \end{figure}
\begin{figure}[h]
  \begin{center}
    \begin{tabular}{cc}
      \resizebox{50mm}{!}{\includegraphics{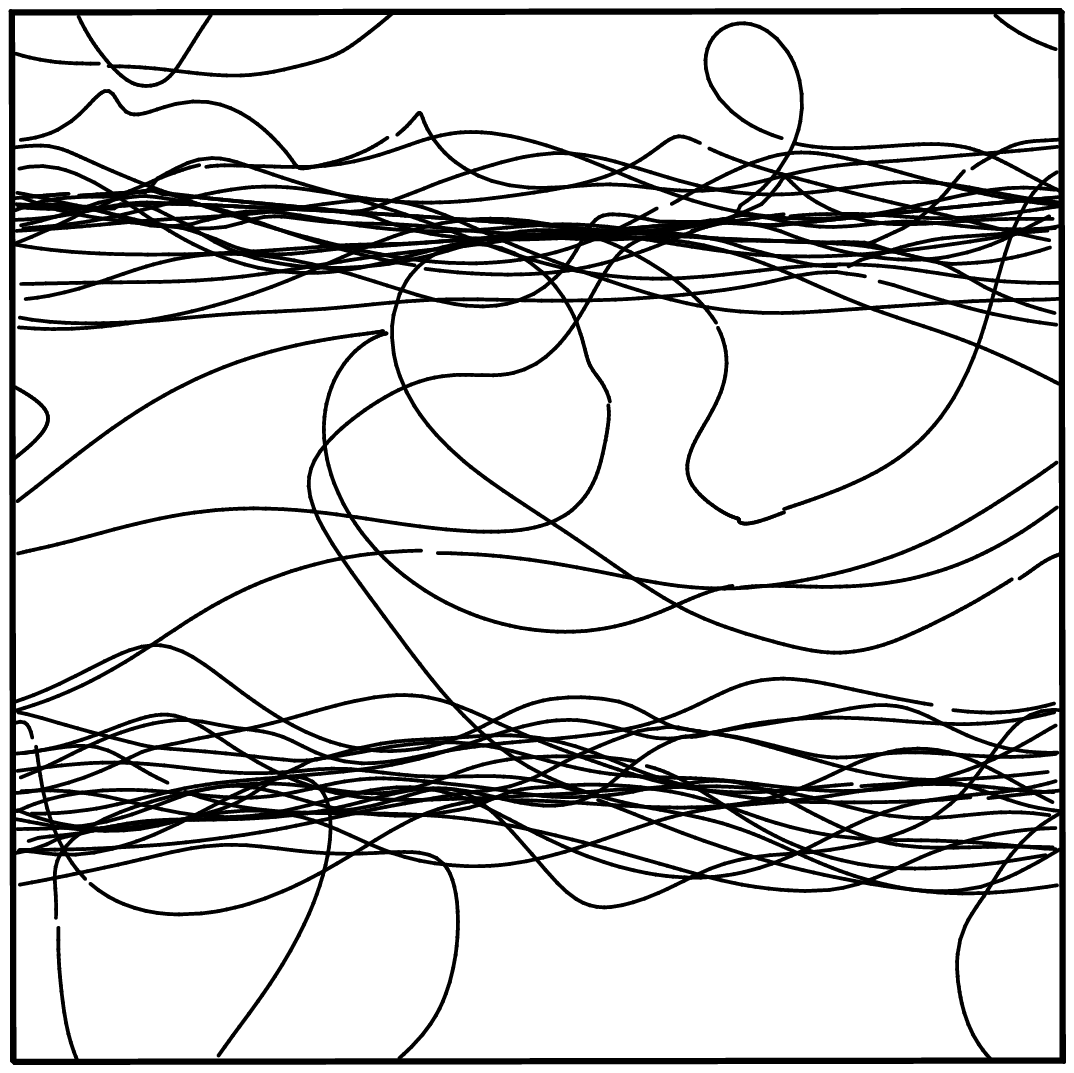}} &
      \resizebox{50mm}{!}{\includegraphics{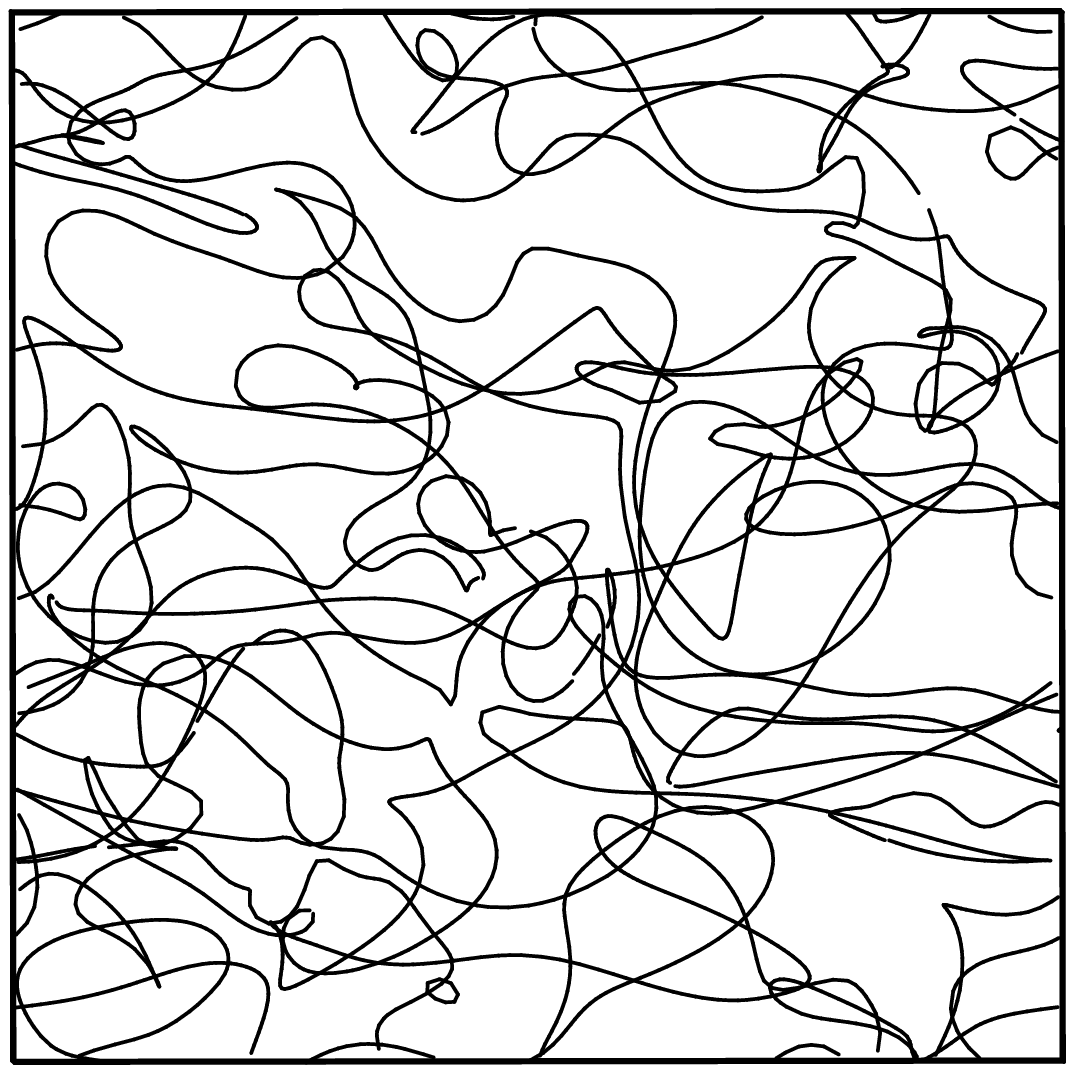}}
    \end{tabular}
     \begin{center}
          \caption{Side view of vortex configuration by the LIA calculation
(left) and by the full Biot-Savart law (right) at t=34.5s. The system is a $(0.2{\rm cm})^3$ cube. Applied normalfluid velocity $v_{ns}=0.367{\rm cm/s}$.}
     \end{center}
    \label{test4}
  \end{center}
\end{figure}
We run both calculations at the temperature $T$=1.6K, in the computing cubic box 0.2 $\times$ 0.2 $\times$ 0.2 ${\rm cm^3}$,
and applied counterflow velocity $v_{ns}$=0.367${\rm cm/s}$.
To explain the difference between the results of the LIA and full Biot-Savart
 law we introduce the dimensionless anisotropy parameter [8]
\begin{equation}
I_{||}=\frac{1}{\Omega L} \displaystyle\int^{}_{\it L}[1-({\bf s}'\cdot{\hat {\bf r}_{||}})^2]d\xi. 
\end{equation}
Here ${\hat {\bf r}_{||}}$ stands for a unit vector parallel to the ${\bf v}_{ns}$ direction, and $\Omega$ is the sample volume. An isotropic tangle yields $\bar{I_{||}}=2/3$. At the other extreme,
if the tangle consists entirely of curves lying in planes normal to ${\bf v}_{ns}$, $\bar{I_{||}}=1$. 
The time evolution of $L(t)$ and $I_{||}(t)$ is shown in Fig 4.
%        \begin{figure}[h]
%          \begin{minipage}{0.5\hsize}
%            \begin{center}
%            \scalebox{0.22}{\includegraphics{T16linedensity.eps}}
%             \label{fig:5a}             
%            \end{center}
%          \end{minipage}
%          \begin{minipage}{0.5\hsize}
%           \begin{center}
%            \scalebox{0.22}{\includegraphics{T16anisotropy.eps}}
%            \label{fig:anisotropy_LIA}
%           \end{center}
%          \end{minipage}
%          \caption{Comparison of a vortex line density $L(t)$ (a) and an 
% anisotropy parameter $I_{||}(t)$ (b).}
%        \end{figure}
\begin{figure}[h]
  \begin{center}
    \begin{tabular}{cc}
      \resizebox{52mm}{!}{\includegraphics{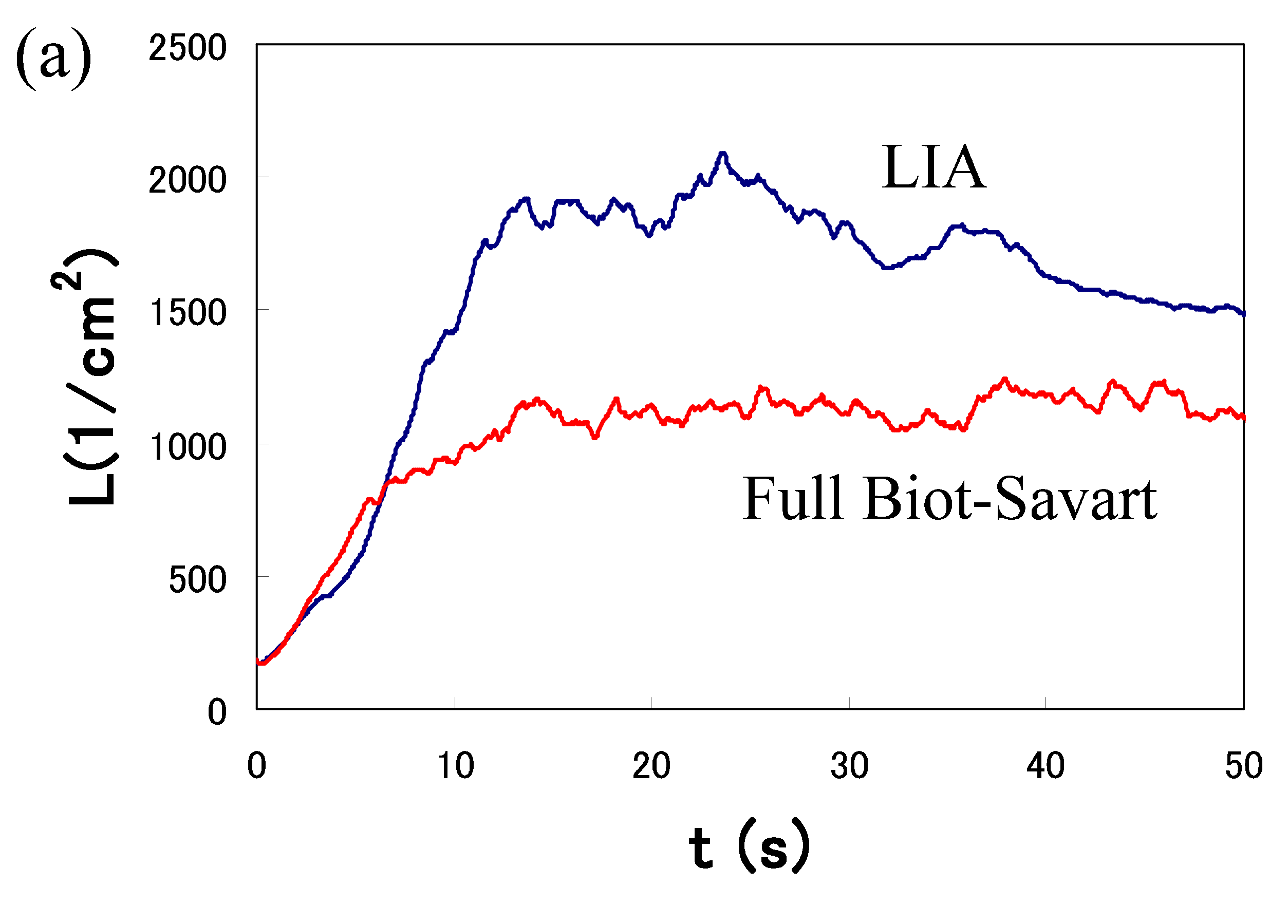}} &
      \resizebox{52mm}{!}{\includegraphics{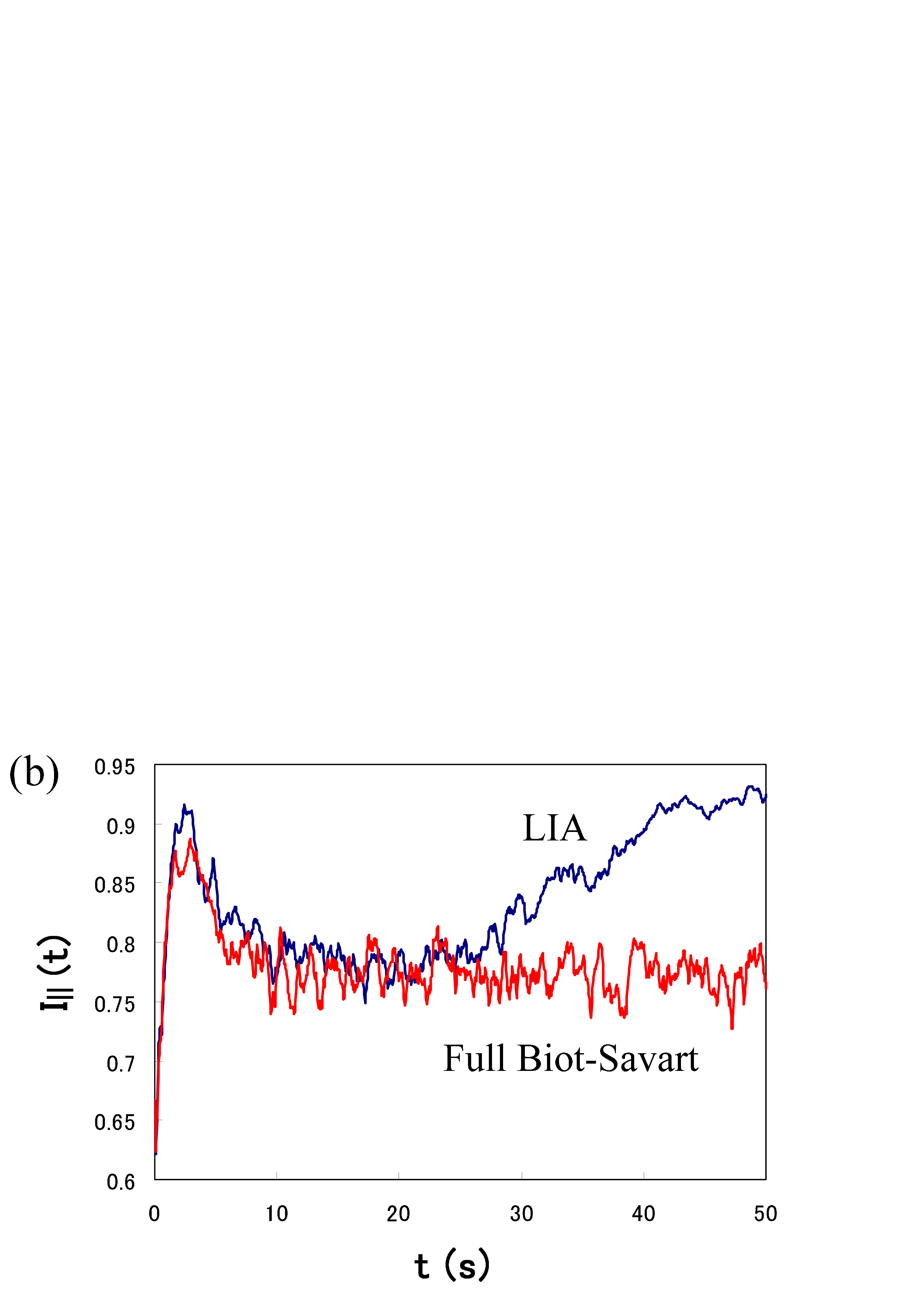}}
    \end{tabular}
     \begin{center}
          \caption{Comparison of a vortex line density $L(t)$ (a) and an 
 anisotropy parameter $I_{||}(t)$ (b).}
     \end{center}
    \label{test4}
  \end{center}
\end{figure}
Figure 3 (left) and Fig.4 (b) show that the many vortices lie in the planes
 normal to
 ${\bf v}_{ns}$, the vortex tangle being degenerate.
Since the dense part of vortices catch other vortices moving freely, 
vortices become to huddle in periodical planes as shown in Fig.3 (left).   
This ill behavior comes from the mutual friction, which tends to expand
 vortices perpendicular to the ${\bf v}_{ns}$, so that vortices gradually lie in planes normal to ${\bf v}_{ns}$. 
However, a non-local term of the Biot-Savart integral could yield the velocity in a direction parallel to ${\bf v}_{ns}$ even when vortices align in a plane
perpendicular to ${\bf v}_{ns}$, thus destroying the ill structure.
Hence, the calculation with the full Biot-Savart law can sustain
 the SSS in contrast to that with the LIA.

\section{Velocity Distribution of Vortices}
For comparison with experimental results [9], we present the velocity
 distributions of vortices in the SSS. 
We obtained the distributions as shown in [Fig.5 left] by measuring the $z$-component of the velocity
 ${\dot{\bf s}}$ at each points on the vortex filaments from the results of 
counterflow turbulence in which applied velocity ${\bf v}_{ns}$ is directed to 
$z$-axis.
The velocity ${\dot{\bf s}}$ does not necessarily express the particle motion
 on the vortex filament which probably occur in the experiments. 
 Therefore we brought this effect in our calculation.
For the sake of simplicity, suppose the trapped particles do not have a 
significant influence on the local motion of the lines, and that viscous
interaction of a particle with the normal fluid causes the particles to move 
along the filament at a rate given by the Stokes law with the force equal to
 the component of the viscous force along the line. With these assumption, 
 the $z$-component of the velocity ${\dot{\bf s}}_p$ including the effect of motion
is derived from the similar fashion of the work [15] like
\begin{equation}
{\dot{\bf s}}_{pz}={\dot{\bf s}}_n\cdot {\hat z}+(v_n-{\dot{\bf s}}_n\cdot
 {\hat z})\cos^2 \theta,
\end{equation}
where ${\dot{\bf s}}_n$ is the normal component of the vortex velocity to the
 vortex filament, $\theta$ is a polar angle of the vortex relative to
 ${\bf v}_{ns}$.
The velocity distributions of both ${\dot{\bf s}}_z$ and ${\dot{\bf s}}_{pz}$
 are shown in [Fig.5 right].
%        \begin{figure}[h]
%          \begin{minipage}{0.5\hsize}
%            \begin{center}
%            \scalebox{0.22}{\includegraphics{velodisT19vn03.eps}}
%             \label{fig:5a}             
%            \end{center}
%          \end{minipage}
%          \begin{minipage}{0.5\hsize}
%           \begin{center}
%            \scalebox{0.22}{\includegraphics{vs_vzT19}}
%            \label{fig:5b}
%           \end{center}
%          \end{minipage}
%          \caption{(a) The vertical velocity $v_z$ distribution at temperature $T$%=1.9K, applied velocity $v_{ns}$=0.572cm/s. The superfluid velocity $v_s$
%is shown by vertical line. (b) Vertical velocity of vortex tangle $\bar{v_z}$, whi%ch is a peak velocity of the distribution, as a function of $v_s$.
%The dashed line corresponds to $\bar{v_z}=v_s$.} 
%        \end{figure}
\begin{figure}[h]
  \begin{center}
    \begin{tabular}{cc}
      \resizebox{52mm}{!}{\includegraphics{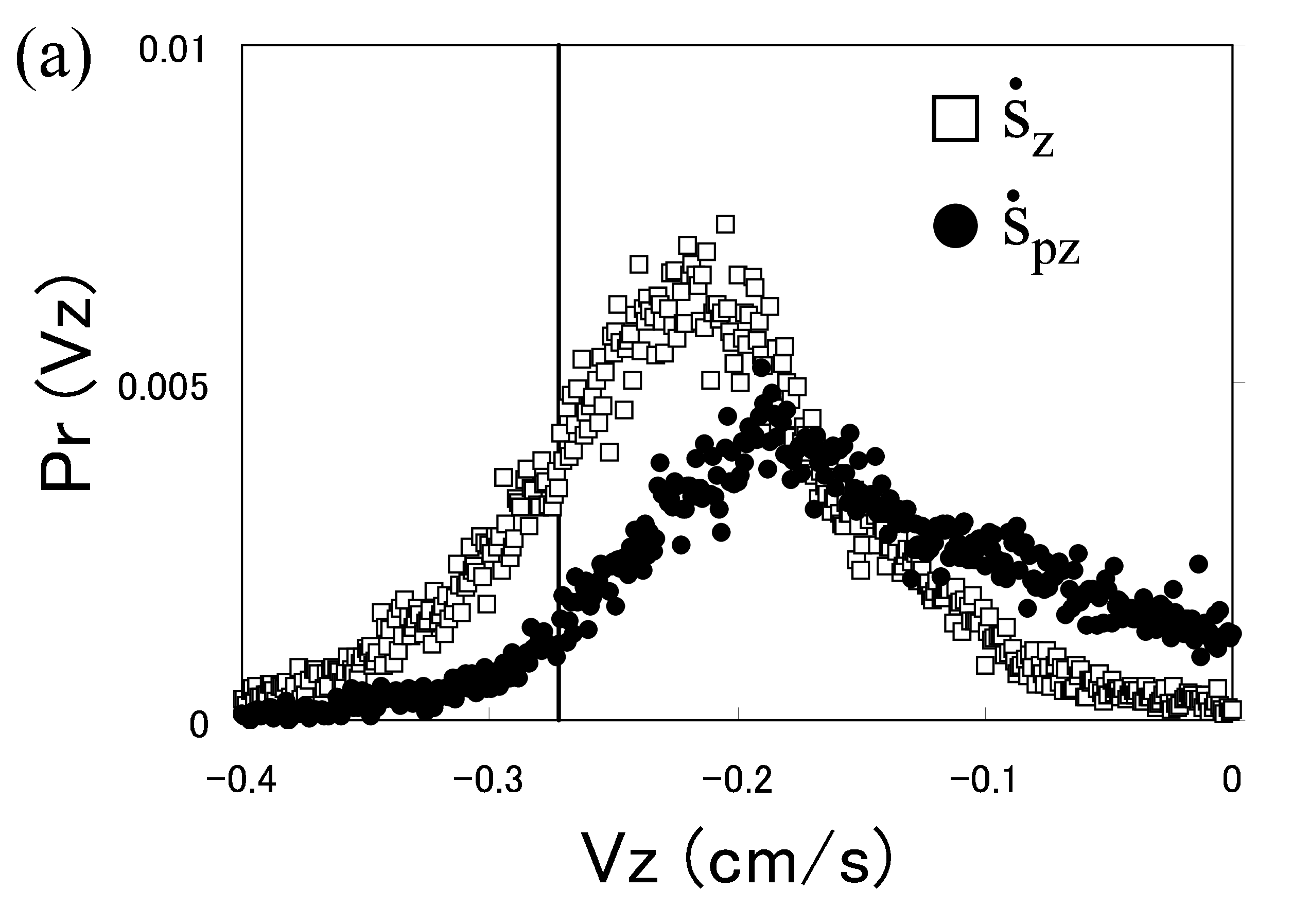}} &
      \resizebox{52mm}{!}{\includegraphics{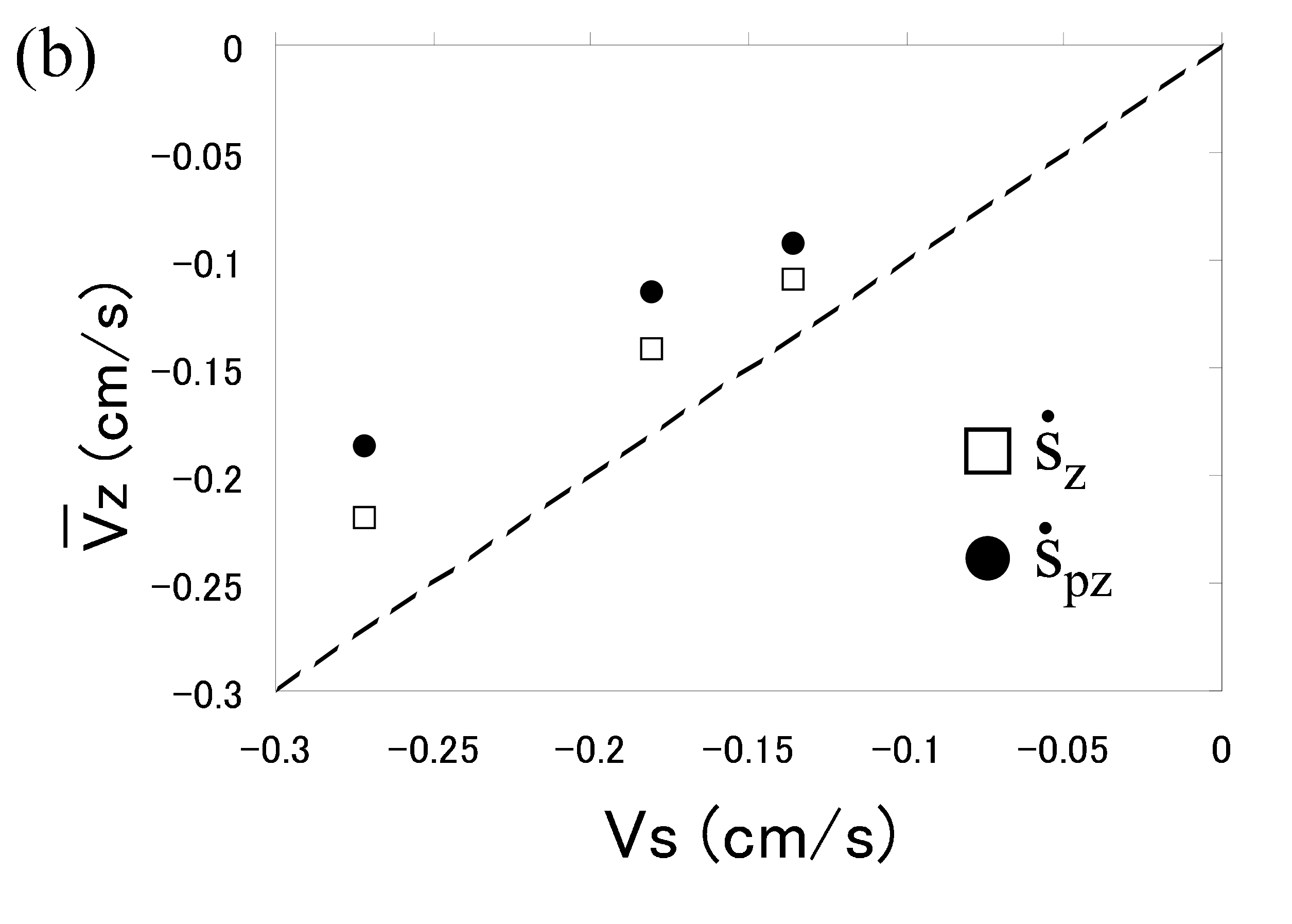}}
    \end{tabular}
     \begin{center}
          \caption{(a) The vertical velocity $v_z$ distribution at temperature $T$=1.9K, applied velocity $v_{ns}$=0.572cm/s. The superfluid velocity $v_s$
is shown by vertical line. (b) Vertical velocity of vortex tangle $\bar{v_z}$, which is a peak velocity of the distribution, as a function of $v_s$.
The dashed line corresponds to $\bar{v_z}=v_s$.}
     \end{center}
    \label{test4}
  \end{center}
\end{figure}
The vortices in our simulations have lower vertical 
velocities than the superfluid velocity just like the experimental
observation.
For the quantitative agreement with observation,
 we need to introduce other effects 
such as a distortion of vortices by particles.

\section{Conclusions}
The obtained numerical results with the full Biot-Savart law demonstrate that
 the initially smooth vortex rings develop to a vortex tangle of 
the statistically steady state.
We compared the numerical results by the full Biot-Savart calculation to
 that by the LIA calculation.
The LIA calculation could not sustain the statistically steady state 
in contrast to the full Biot-Savart calculation.
 We compared the velocity distributions of vortices obtained by our numerical simulations to the observations by the visualization experiment of couterflow by using solid hydrogen particles.
Our results agree with the experimental observation qualitatively in that most vortices have lower velocities than the superfluid velocity.

\begin{acknowledgements}
We would like to thank M. S. Paoletti and D. P. Lathrop for giving their preliminary experimental data and the fruitful discussions.
\end{acknowledgements}

%\begin{figure}
%  \includegraphics{example.eps}
%\end{figure}
%\begin{figure*}
%  \includegraphics[width=0.75\textwidth]{example.eps}
%\end{figure*}
%\begin{table}
%\begin{tabular}{lll}
%\hline\noalign{\smallskip}
%\end{tabular}
%\end{table}

%\begin{acknowledgements}
%\end{acknowledgements}



\end{document}